\documentclass[journal,draftcls,onecolumn,12pt,twoside]{IEEEtranTCOM}
\usepackage{url}
\usepackage[dvips]{color}
\usepackage{graphicx}
\usepackage{graphics}
\usepackage{latexsym}
\usepackage{showidx}
\usepackage{latexsym}
\usepackage{amssymb}
\usepackage{amsfonts}
\usepackage{amsmath}
\usepackage{amssymb}
\usepackage{amsfonts}
\usepackage{amsmath}
\usepackage{mathrsfs}

\newcommand{\nv}{\mathbf{n}}
\newcommand{\rv}{\mathbf{r}}
\newcommand{\sv}{\mathbf{s}}

\newcommand{\xv}{\mathbf{x}}
\newcommand{\yv}{\mathbf{y}}

\newcommand{\CC}{\mathbb{C}}
\newcommand{\RR}{\mathbb{R}}
\newcommand{\ZZ}{\mathbb{Z}}

\newcommand{\Cc}{\mathcal{C}}

\begin{document}

\title{Performance of Lattice Coset Codes \\ on Universal Software Radio Peripherals}
\author{\IEEEauthorblockN{Jinlong Lu, J.~Harshan, and Fr\'ed\'erique Oggier
\thanks{
This work was funded in part by the Singapore National Research Foundation under Research Grant NRF-RF2009-07 and under Research grant NRF/2014/MDT14.
}
}\\
\IEEEauthorblockA{Division of Mathematical Sciences,\\
School of Physical and Mathematical Sciences,\\
Nanyang Technological University, Singapore}\\
{\small {\tt $\{$kerin\_lu, jharshan, frederique$\}$@ntu.edu.sg}}}
\and

\maketitle
\begin{abstract}
We consider an experimental setup of three Universal Software Radio Peripherals (USRPs) that implement a wiretap channel, 
two USRPs are the legitimate players Alice and Bob, while the third USRP is the eavesdropper, whose position we vary to evaluate information leakage. The experimented channels are close to slow fading channels, and coset coding of lattice constellations is used for transmission, allowing to introduce controlled randomness at the transmitter. Simulation and measurement results show to which extent coset coding can provide confidentiality, as a function of Eve's position, and the amount of randomness used.
\end{abstract}
\begin{center}
\end{center}

\section{Introduction}
\label{sec:intro}

We consider a wiretap channel, comprising a legitimate transmitter, Alice, and two receivers: a legitimate one, Bob, and a passive adversary, Eve. For the legitimate users Alice and Bob, both reliable and confidential transmission needs to be achieved, while Eve is trying to eavesdrop the communication. This is done through wiretap coding. Alice encodes her secret $\sv$ into a codeword $\xv$ belonging to a code $\Cc$, and $\xv$ is then sent through the wiretap channel to Bob and Eve, which respectively receives $\yv_B$ and $\yv_E$. What distinguishes wiretap coding from standard coding is the constraint on confidentiality, which should be obtained without invoking cryptographic means: confidentiality is obtained by a suitable injection of controlled randomness mixed with an appropriate coding strategy at the transmitter, which enables Bob to receive his message with high probability, while confusing the eavesdropper to the point of making her knowledge of the secret message negligible. This is formally expressed
by saying that the mutual information between what Eve receives and the secret is zero:
\begin{equation}\label{eq:strongsec}
I(\sv;\yv_E) = H(\sv)-H(\sv|\yv_E)=0
\end{equation}
or equivalently, that the entropy $H(\sv|\yv_E)$ of the secret knowing the received message at the eavesdropper is the same as the entropy of the secret.

Wiretap coding necessarily requires the channel from Alice to Bob to be different from that from Alice to Eve. What ``different" means, as well as wiretap coding strategies, depend on the channel model, e.g., discrete memoryless, additive white Gaussian, Rayleigh fading, or multiple-inout multiple output (MIMO), to name a few popular models. We refer the readers to \cite{LP09} and \cite{LO13} for a survey of information theoretic results and respective coding strategies for wiretap channels, and to \cite{KMB}-\cite{ZLGYY} for application of practical codes on wiretap channels.

In all cases, the channel assumptions, in particular regarding the eavesdropper's channel, are critical, since the confidentiality analysis relies on them.
This is the case for wiretap coding, but also for any other schemes whose security relies on channel noise, such as secret key generation.

The goal of this paper is to study wiretap coding from an experimental view point using a USRP testbed comprising three USRPs, one for each of the three players, Alice, Bob and Eve. The channels between Alice and Bob, and Alice and Eve respectively, are close to slow fading channels, whose SNRs and noise are given by the experimental settings. Transmission is done using signal constellations from lattices, which are transmitted using coset coding as explained in Section \ref{sec:coset}, to introduce controlled randomness (for that reason, we will use the term ``coset coding" rather than wiretap coding in the rest of the paper). The positions of Alice and Bob are kept fixed, while we vary the position of Eve to analyze her received signal, and how much information is leaked depending on both her position and the coding scheme used.

We present both simulations and experimental results that consistently show how coset coding does provide confusion at the eavesdropper, with entropy (see Subsection \ref{subsec:ce}) and decoding error as metrics, using an optimal decoder for Eve, as proven in Subsection \ref{sec:sod}. Extensive results are provided in Section \ref{sec:simul} and Section \ref{sec:exp_results}, to compare coset coding versus conventional coding, but also different coding schemes using lattice constellation of different dimensions, different amounts of randomness and different positions for Eve. Experiments were realized by transmitting ``the cameraman image" (see Fig.~\ref{fig:cameraman}), which furthermore allows a visualization of the effect of coset coding (see Fig.~\ref{fig:camera_man_rx}).

We believe that this type of experimental work is critical to the development of physical layer security, since it gives an insight of how practical a security scheme such as coset coding behaves in practice, without having to rely on channel assumptions.
Another work with the same philosophy was done for key generation in \cite{PCB}, 
where the authors investigate the role of the eavesdropper's statistics when actually implementing a secret-key generation system over a wireless channel. This is done via a software-defined radio testbed, where the channel gains are measured. The experimental setup shows a 20\% loss in secret-key rate with respect to theoretical bounds.  

\section{Coset Encoding of Lattice Codes}
\label{sec:coset}

\begin{figure}
\begin{center}
\includegraphics[scale=0.4]{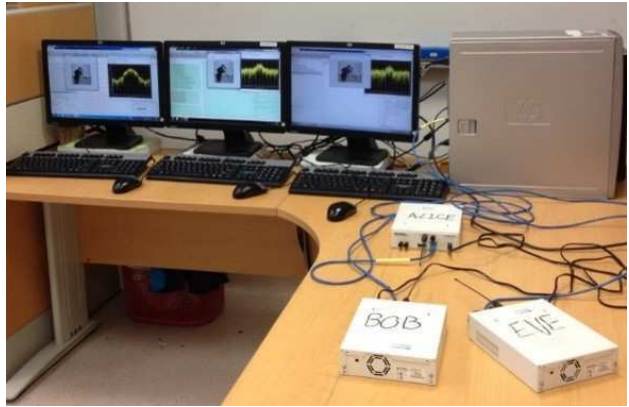}
\caption{\label{fig:setup}
USRP testbed, comprising three USRPs: a transmitter (Alice) and two receivers
(Bob and Eve).
}
\end{center}
\end{figure}

We consider a wiretap testbed formed by three USRPs, as shown on Fig. \ref{fig:setup}. One USRP plays the role of the legitimate transmitter Alice, while the other two USRPs are the receivers, Bob and Eve.

This wiretap channel is modeled by
\begin{eqnarray}
\yv_B & = & h_B\xv + \nv_B \label{eq:channelbob} \\
\yv_E & = & h_E\xv + \nv_E \label{eq:channeleve}
\end{eqnarray}
where $h_B,h_E \in \CC$ are the respective channel gains, 
$\xv \in \CC^{L/2}$ is the transmitted message ($L$ is the real dimension, an even number),
and $\nv_B,\nv_E$ are the respective channel noises at Bob and Eve, distributed as circularly symmetric complex Gaussian, denoted by 
as $\Cc\mathcal{N}({\bf 0},\sigma^2 {\bf I}_{L/2})$.

To transmit over this complex channel, we consider lattice coding.
A real lattice $\Lambda$ of dimension $L$ is a discrete set of points in $\RR^L$ generated as integral linear combinations of a set of $L$ linearly independent vectors in $\RR^L$. 
For actual data transmission, a finite constellation of the lattice is chosen. For example, QAM constellations are obtained by taking a finite subset of the lattice $\ZZ^L$.
When $L$ is even, a real lattice can be used for transmission over a complex channel of dimension $L/2$.
The role of a lattice encoder is to map a bit string to a lattice point. 
However for wiretap coding, we use instead lattice coset coding, which allows us to introduce randomness.

\subsection{Lattice Coset Coding}
\label{subsec:lcd}

\begin{figure}
\begin{center}
\includegraphics[scale=0.3]{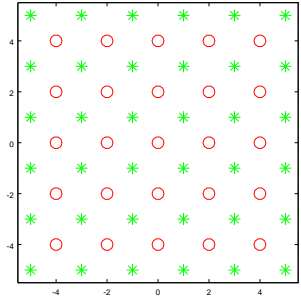}
\caption{\label{fig:D2} The lattice $D_2$ obtained via Construction A.
The circles are used for the lattice $2\ZZ^2$, and the stars for the coset $2\ZZ^2+(1,1)$.
}
\end{center}
\end{figure}

At the transmitter, we implement a lattice coset encoder. 
A lattice coset encoder requires two nested lattices $\Lambda_E \subset \Lambda_B$, and a partition of $\Lambda_B$ as a union of cosets of $\Lambda_E$:
\begin{equation}\label{eq:cosetlatt}
\Lambda_B = \cup_{\sv\in \Lambda_B/\Lambda_E} (\sv+\Lambda_E)
\end{equation}
where $\sv$ is a coset representative. In Fig. \ref{fig:D2}, a lattice $\Lambda_B$ is shown as the union of the lattice $\Lambda_E = 2\ZZ^2$ and its coset $2\ZZ^2+(1,1)$.

In the context of wiretap coding, coset coding is used to create confusion by introducing controlled randomness. In this case, $\sv$ actually encodes the secret, while a vector $\rv$ is chosen randomly inside $\Lambda_E$, to obtain $\xv = \sv + \rv$. We use the index $B$ in $\Lambda_B$ to indicate that this lattice is used for transmission to Bob, while $\Lambda_E$ is the lattice meant to create confusion at Eve's end.

A partition of $\Lambda_B$ as a union of cosets of a sublattice $\Lambda_E$ can be obtained using the so-called Construction A \cite{splag}. Let $\rho:\ZZ^L \rightarrow \{0,1\}^L$ denote the componentwise reduction modulo 2, which maps an integral vector to a binary one. In $\{0,1\}^L$, choose a binary linear code $C$ of length $L$ and dimension $k$. Then we take for $\Lambda_B$ the lattice $\rho^{-1}(C)=2\ZZ^L+C$ (possibly scaled), and thus $\Lambda_E=2\ZZ^L$. Codewords from $C$ are the coset representatives. The lattice on Fig. \ref{fig:D2} is called $D_2$, and is actually obtained using Construction A and the repetition code $C=\{(0,0),(1,1)\}$:
\[
D_2=2\ZZ^2+C=(2\ZZ^2+(0,0))\cup(2\ZZ^2+(1,1)).
\]

Table \ref{tab:constA} lists different constructions A of lattices that we use.

\begin{table}
\begin{center}
\begin{tabular}{|l|l|l|}
\hline
$\Lambda$ & $L$ & $C$ \\
\hline
$\ZZ^2$ & $2$ & $\{00,01,10,11\}$\\
$D_2$   & $2$ & $\{00,11\}$\\
\hline
$D_4$   & $4$  & $(4,3,2)$ parity check \\
\hline
$\sqrt{2}E_8$ & $8$ & $(8,4,4)$ Reed-M\"uller\\
\hline
\end{tabular}
\caption{\label{tab:constA}
Lattices via Construction A: $\Lambda=2\ZZ^L+C$.}
\end{center}
\end{table}

For an infinite lattice, the vector $\rv$ introduced for randomness is chosen according to the uniform distribution. To perform experiments, finite constellations are carved from lattices, by taking lattice points in the hypercube $\{0,1,2,\ldots,M-1\}^L$, and we keep the choice of a uniform distribution for the randomness used (a possibility for further experiments could be to use instead the distribution proposed in \cite{LLBS14}).

In what follows, we use the terminology {\em standard} or {\em conventional encoding} to refer to lattice coding (say data points are mapped to points in $D_2$ or $E_8$) without use of randomness, in contrast to {\em coset encoding} which introduces randomness as explained above.

\subsection{Conditional Entropy at Eve}
\label{subsec:ce}

Recall from (\ref{eq:strongsec}) that confidentiality is measured by
\[
I(\sv;\yv_E) = H(\sv)-H(\sv|\yv_E),
\]
and coset encoding should ensure that $I(\sv;\yv_E)=0$.
Ideally, we thus would want $H(\sv|\yv_{E}) = H(\sv)$. With this objective in mind, it is interesting to observe how close are the values of $H(\sv|\yv_{E})$ and $H(\sv)$, that arise from the coset encoding schemes in practice (to meet the confidentiality criterion). Since conditional entropy is independent of any specific decoder implementation at Eve, we use it to showcase the impact of coset encoding for creating confusion at Eve. To get a sense of the conditional entropy of coset encoding, we conducted Monte-Carlo simulations assuming a fixed channel gain $h$, for the following one-dimensional ($L=1$) encoding schemes:
\begin{itemize}
\item Coset encoding using $\Lambda_{E} = 2\mathbb{Z}$ and $\Lambda_{B} = \mathbb{Z}$, where 1 bit of secret is mapped to a coset in $\mathbb{Z}/2\mathbb{Z}$, and the coset representative is chosen with 2 bits of randomness. If the coset $2\mathbb{Z}$ is chosen, then one of the elements in $\{0, 2, 4, 6\}$ is uniformly chosen as the representative. Similarly, one of the elements in $\{1, 3, 5, 7\}$ is uniformly chosen as the coset representative for the coset $2\mathbb{Z} + 1$.
\item Conventional encoding using $\mathbb{Z}/2\mathbb{Z}$, where 1 bit of secret is deterministically mapped to the coset representatives $\mathbb{Z}/2\mathbb{Z} = \{0, 1\}$ (no randomness added).
\item Conventional encoding using $\mathbb{Z}/8\mathbb{Z}$, where 3 bits of secret is deterministically mapped to the coset representatives $\mathbb{Z}/8\mathbb{Z} = \{0, 1, 2, \ldots, 7\}$ (no randomness added).
\end{itemize}
To empirically compute conditional entropy, we transmitted a random ensemble of points from the above encoding schemes (after suitably scaling and shifting to keep their average transmit powers identical). Upon receiving the points $\yv_{E}$ in \eqref{eq:channeleve}, we evaluated the conditional probability mass function $P(\sv | \yv_{E} = y)$, by looking at various $\sv$ values that result in a particular realization of $\yv_{E}$, say $\yv_{E} = y$. Then the conditional probability mass function on $\sv$ was obtained using frequency of occurrence of various realizations of $\sv$. Note that experimentally, obtaining \emph{multiple} realizations of $\sv$ that result in the same real number $y$ is unlikely. As a result, we partitioned the real line into tiny bins so as to collect multiple realizations falling within the same bin. This allowed us to compute

\begin{figure}
\begin{center}
\includegraphics[scale = 0.25]{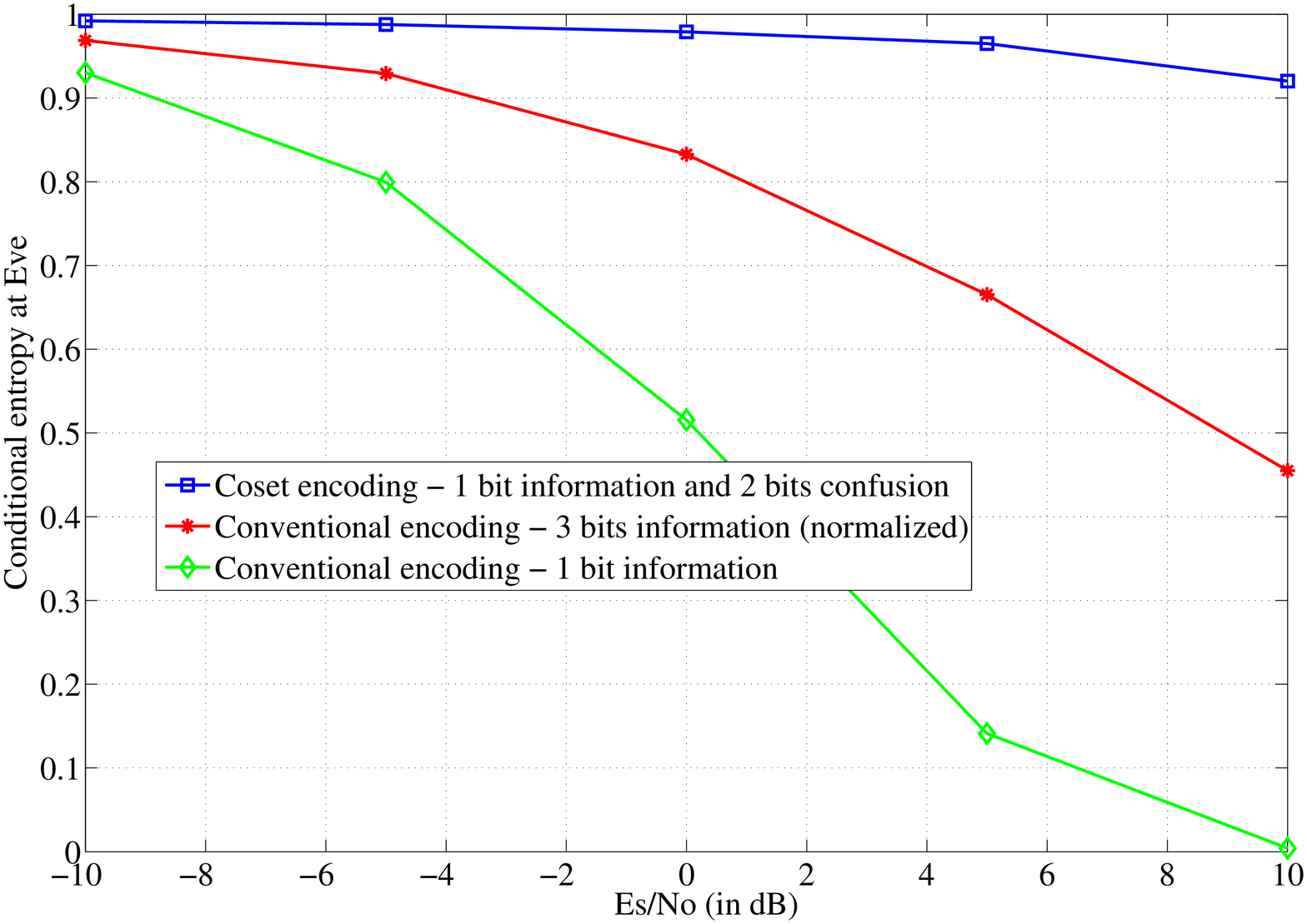}
\vspace{-0.5cm}
\caption{\label{fig:entr_coset_vs_conv} Conditional entropy of various encoding schemes when measured at Eve. For the conventional scheme with 3 bits of information, conditional entropy is normalized to 1 bit for comparison purposes with other schemes. The plots show that coset encoding with 1 bit information and 2 bits of randomness results in conditional entropy close to 1 at Eve, thereby serving the confidentiality purpose.}
\end{center}
\end{figure}

\begin{small}
\[
H(\sv| \yv_{E} = y) = -\sum_{j} P(\sv = j| \yv_{E} = y)\log_{2} P(\sv = j | \yv_{E} = y),
\]
\end{small}

\noindent where $j$ is used to denote different realizations of the secret $\sv$. From  the above expression, the conditional entropy numbers are obtained as
\[
\tilde{H}(\sv | \yv_{E}) = -\sum_{y} P(y)H(\sv | \yv_{E} = y),
\]
where $\tilde{H}(\sv | \yv_{E})$ is an approximation to the standard entropy definition.  The measured conditional entropy values for the three encoding schemes are given in Fig. \ref{fig:entr_coset_vs_conv}. The plots show that the conditional entropy of coset encoding is close to 1 (which is the entropy of the secret), much higher than the other two conventional encoding schemes. 

\subsection{Suboptimal and Optimal Decoders}
\label{sec:sod}

In the previous subsection, we studied the conditional entropy criterion, which is independent of the decoder employed at Eve. We next discuss relevant decoders for the coset encoding scheme. Consider a lattice code $\mathcal{C} \subset \mathbb{C}^{L/2}$ which can be partitioned into $2^{k}$ equally sized subsets as follows 
\begin{equation*}
\mathcal{C} = \mathcal{C}_{0} \cup \mathcal{C}_{1} \cup \ldots \cup \mathcal{C}_{2^{k}-1},
\end{equation*}
where $|\mathcal{C}_{0}|=\ldots = |\mathcal{C}_{2^k-1}|$ is a power of 2. At the encoder, based on $k$ information bits, or rather $k$ bits forming the secret $\sv$ in our case, one of the sets in $\{ \mathcal{C}_{0}, \mathcal{C}_{1}, \ldots, \mathcal{C}_{2^{k}-1} \}$ is chosen, and then a random codeword $\textbf{x}$ from the chosen set is transmitted. The choice of the random codeword is based on $\log_{2}(|\Cc_0|)$ random bits that are intended to create confusion at Eve. Specifically, if the $l$-th codeword in $\mathcal{C}_{j}$ is chosen, for some $0 \leq j \leq 2^{k} - 1$, then $\textbf{x}$ will take the codeword denoted by $\textbf{c}^{(l)}_{j}$. With that, the signal received at Eve is given by (\ref{eq:channeleve}), namely (we drop the subscript $E$ for the rest of this subsection to lighten the notation)
\begin{equation*}
\textbf{y} = h \textbf{x} + \textbf{n},
\end{equation*}
where $h$ is the fading gain, $\textbf{n}$ is the AWGN distributed as circularly symmetric complex Gaussian, denoted by $\mathcal{CN}(\mathbf{0}, \sigma^{2}\mathbf{I}_{L/2})$. When the average transmit power per channel use is $P$, the average Signal-to-noise ratio (SNR) is given by $\rho \triangleq \frac{P |h|^2}{\sigma^{2}}$.

The $k$-length bit sequence $\hat{\textbf{s}}$ can be ML decoded as
\begin{eqnarray*}
\hat{\textbf{s}} & = & \arg \max_{\{\mathcal{C}_{j}\}} P(\textbf{x} \in \mathcal{C}_{j} | \textbf{y}, h),
\end{eqnarray*}
where $P(\cdot)$ denotes the conditional probability density function. By applying Bayes' rule, we have
\begin{eqnarray}
\hat{\textbf{s}} 
& = & \arg \max_{\{\mathcal{C}_{j}\}} \frac{P(\textbf{y} | h, \textbf{x} \in \mathcal{C}_{j})P(\textbf{x} \in \mathcal{C}_{j} | h)}{P(\textbf{y} | h)} \nonumber \\
& = & \arg \max_{\{\mathcal{C}_{j}\}} \sum_{l = 1}^{|\Cc_j|} P(\textbf{y} | \textbf{x} = \mathbf{c}^{(l)}_{j}, h) \nonumber \\
& = & \arg \max_{\{\mathcal{C}_{j}\}} \sum_{l = 1}^{|\Cc_j|} e^{-\rho||\textbf{y} - h\textbf{c}^{(l)}_{j}||^{2}}.
\label{eq:ML}
\end{eqnarray}
The second equality follows since $P(\textbf{y} | h)$ is fixed and $P(\textbf{x} \in \mathcal{C}_{j} | h)$ is a constant. The last equality follows from the fact that $\textbf{y}$ is Gaussian distributed when conditioned on $\textbf{x}$ and $h$. At high $\rho$ values, \eqref{eq:ML} can be approximated as 
\begin{eqnarray}
\hat{\textbf{s}} & = & \arg \max_{\{\mathcal{C}_{j}\}} e^{-\rho||\textbf{y} - h \textbf{c}^{(l')}_{j}||^{2}} \nonumber \\
& = & \arg \min_{\{\mathcal{C}_{j}\}} ||\textbf{y} - \textbf{c}^{(l')}_{j}||^{2},
\label{eq:MD}
\end{eqnarray}
where $\textbf{c}^{(l')}_{j}$ is the closest codeword to $\textbf{y}$ (in Euclidean distance) among the members of $\mathcal{C}_{j}$. We refer to \eqref{eq:ML} and \eqref{eq:MD} as the Maximum Likelihood (ML) decoder and Minimum Distance (MD) decoder, respectively. In our preliminary work in \cite{itw}, we have used the MD decoder to obtain experiment results on the error performance of coset encoding scheme. However, in this work, we employ the superior ML decoder to obtain the results. The use of \eqref{eq:ML} is imperative to showcase that despite the optimality of ML decoder, lattice code based coset encoding scheme results in higher confusion at Eve than the other conventional schemes. 


\begin{figure}
\begin{center}
\includegraphics[scale = 0.7]{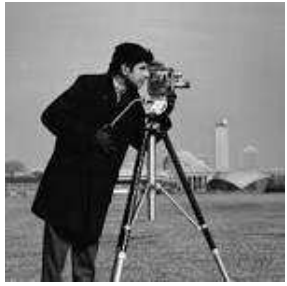}
\caption{\label{fig:cameraman} The \emph{Cameraman} image transmitted from Alice to Bob. This image is extensively used as a test
image in image processing research. This \emph{Cameraman} image is transmitted in an unencrypted form so as to validate the effect of wiretap coset encoding.}
\end{center}
\end{figure}

Having introduced the optimal decoder for the coset encoding schemes, in the next section we use them to showcase how much secret can be extracted at Eve from $\yv_{E}$.

\section{Experiment Set up using USRPs}
\label{sec:simul}

We illustrate the role and effectiveness of coset coding through experiments over a testbed, comprising National Instrument USRP-2920 devices (as shown on Fig. \ref{fig:setup}) to implement the wiretap channel consisting of Alice, Bob and Eve (all are single antenna devices). The experiments were carried out by fixing the position of Alice and Bob while placing Eve at different positions. In particular, Bob is positioned at some fixed distance from Alice, referred to as Placement 1, whereas Eve is placed at five different distances, referred to as Placement 1, 2, 3, 4 and 5 (in the order of increasing distance).\footnote{The use of 5 placements 1, 2, 3, 4 and 5 are only for the sake of introducing the setup. However, placements other than the above 5 have also been used to generate experiment results.} Alice transmits the black-and-white image shown in Fig. \ref{fig:cameraman} using either
conventional coding or coset coding, whereas Eve attempts to recover this image at different positions. Using an image for the experiments gives a visualization of the effect of coset coding through the quality of the reconstructed image at Eve, i.e., more
distortion of the recovered image implies more confusion at Eve. Since we have already discussed the benefits of coset encoding in terms of conditional entropy, we now present experiment results based on decoding error.

\begin{center}
\begin{table}
\caption{\label{table_syst_parameters} System parameters for the experiments. SDR denotes software-defined radio, and LO denotes local oscillator.}
\begin{center}
\begin{small}
\begin{tabular}{|c|c|c|c|c|c|c|c|c|c|c|}
\hline Parameter & SDRu Transmitter & SDRu Receiver\\
\hline Center Frequency (Hz) & 2.42E+09 & 2.42E+09\\
\hline LO offset(Hz) & 0 & 0\\
\hline Gain(dB) & 18  &  31\\
\hline Interpolation/ Decimation & 500 & 500\\
\hline
\end{tabular}
\end{small}
\end{center}
\end{table}
\end{center}
\begin{center}
\begin{table}
\caption{\label{table_RCF} Parameters for Rx/Tx Raised Cosine filter}
\begin{center}
\begin{small}
\begin{tabular}{|c|c|c|c|c|c|c|c|c|c|c|}
\hline Parameter & SDRu Transmitter/Receiver\\
\hline Filter shape & square root\\
\hline Rolloff Factor & 0.5\\
\hline Filter span in symbols & 6\\
\hline Input/Output samples per symbol & 4\\
\hline linear amplitude filter gain & 1\\
\hline
\end{tabular}
\end{small}
\end{center}
\end{table}
\end{center}

\begin{figure}
\begin{center}
\includegraphics[scale = 0.5]{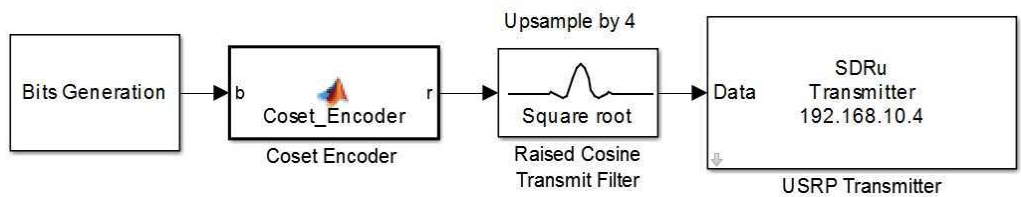}
\caption{\label{tx_coset_encoder_block} Transmitter side baseband operations as implemented in our experiments on Matlab.}
\end{center}
\end{figure}

The system parameters of the experiment setup are listed in Table \ref{table_syst_parameters} and \ref{table_RCF}. More details on the setup are available in \cite{USRP_webpage}. Communication between the USRP devices takes place over a sequence of frames where each frame constitutes $n=100$ complex symbols in baseband representation. Among the 100 symbols, the first 13 symbols are QPSK modulated pilot symbols which are constructed by identically placing a 13-length binary Barker code sequence \cite{barker_code_cite} on the in-phase and quadrature components. These pilot symbols are essential for the timing recovery and frame synchronization operations at the receiver. The rest of the 87 locations in the frame are allocated for data transmission. On the application level, the digital
image is first broken down into a sequence of binary digits (around 64K bytes in size) and then these bits are communicated to Bob by spreading them over a sequence of frames. For the uncoded scheme, 87 binary symbols (from the set $\{0, 1\}$) are embedded into the frame, while for the coded schemes (both conventional and coset), these locations are loaded with uniformly chosen codewords from an underlying lattice code (carved from $\Lambda_B$, with $\Lambda_B \in \{D_{2},D_{4},E_{8}\}$). For the coded schemes, whenever the codewords do not fill the 87 symbols completely, the remainder locations are loaded with dummy symbols. Further, note that the codewords of the chosen lattice codes in $\mathbb{Z}, D_{2}$, $D_{4}$ and $E_{8}$ have components with only non-negative real values. Hence, for implementation purpose, the symbols are appropriately shifted around the origin to reduce the average transmit power.  

Our method to transmit the lattice codewords of $\mathbb{Z}$, $D_{2}$, $D_{4}$ and $E_{8}$ is as follows. Irrespectively of whether the strategy is conventional encoding or coset encoding, for any $\textbf{x} \in \{0, 1, \ldots, M-1\}^{L}$, where $L$ is the dimension of the lattice, the transmitted codeword is of the form 
\begin{equation}
\label{off_set}
\textbf{x}_{t} = \frac{e^{-i \frac{\pi}{4}}}{\sqrt{E_{avg}}}\left(\textbf{x} - \frac{M-1}{2} \right) \in \mathbb{C}^{L},
\end{equation}
where $E_{avg}$ is the average energy of the shifted constellation $\{-\frac{M-1}{2}, -\frac{M-1}{2} + 1, \ldots, \frac{M-1}{2} - 1, \frac{M-1}{2}\}$. Using the scale and the shift operation in \eqref{off_set}, each component of $\textbf{x}_{t}$ takes value from a complex constellation with unit average energy. For the experiment, we use the carrier frequency compensation and timing recovery blocks that are tailor-made for QPSK signal sets \cite{mathworks}. Hence, to facilitate this reuse, the real lattice points are rotated by $-45$ degrees.

Once a frame constituting Barker code symbols and lattice codewords $\{\textbf{x}_{t}\}$ is generated, it is subsequently upsampled by $4$ and then passed through a square-root raised cosine filter (with roll-off factor 0.5) for pulse shaping purpose (see Fig. \ref{tx_coset_encoder_block} for the transmitter side baseband operations). Finally, the filtered samples are forwarded to the USRP device (with interpolation factor 500) for passband transmission.

\subsection{Reception and Decoding}
\label{subsec:RaD}

\begin{figure}
\begin{center}
\includegraphics[scale=0.4]{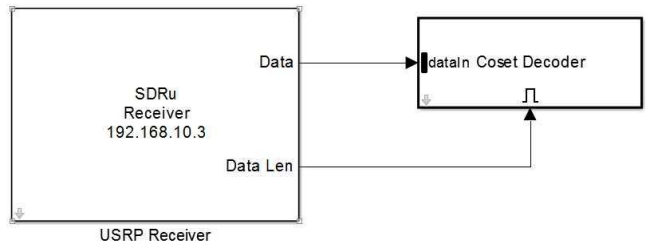}
\end{center}
\caption{\label{rx_block} Receiver block as implemented in our experiments on Matlab.}
\end{figure}

\begin{figure*}
\begin{center}
\includegraphics[scale=0.4]{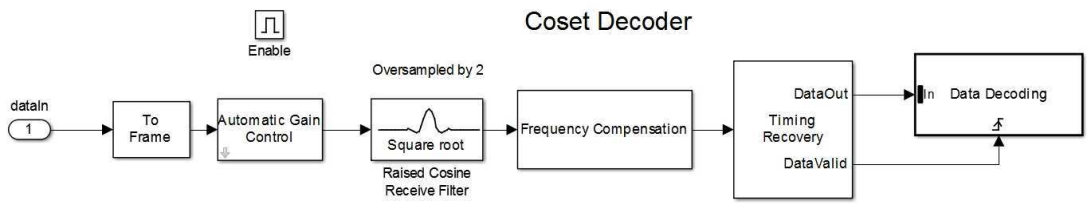}
\end{center}
\caption{\label{coset_decoder_block} Receiver side baseband operations as implemented in our experiments on Matlab.}
\end{figure*}

At the receiver, the passband signal is down-converted and appropriately sampled by the USRP hardware. Then these samples are forwarded to the simulink block (with decimation factor 500) for baseband processing (see Fig. \ref{rx_block} and \ref{coset_decoder_block} for the receiver side baseband operations).

A total of 4000 samples (equivalent to 10 received frames) are processed batch-wise in order to facilitate frame synchronization and timing recovery operations. For accurate synchronization, the received samples are passed through an Automatic Gain Control (AGC) block before it is filtered using a square-root raised cosine receive filter (with identical parameters as that of the transmit filter). Subsequently, the filtered output is forwarded to the frequency compensation and timing recovery blocks. In the timing recovery block, the Barker code sequence is independently generated, using which the beginning of the received frames is detected through cross-correlation operations. Finally, the \emph{aligned} frames are forwarded to the data decoding block (either conventional or coset), which assuming the frame synchronization is accurate, estimates the channel gain using the Barker code symbols on the first 13 locations. Since the AGC gain values may vary across successive frames, we use the estimated channel gain locally within a frame but not across frames. The received lattice codewords by Eve of the form $$\mathbf{y}_E = h_E\textbf{x}_{t} + \mathbf{n}_{E}$$ are extracted from the frame to carry out the decoding process. Here, the scalar $h_E$ is the complex channel gain (for Eve) and $L \in \{ 1, 2, 4, 8\}$ is the block length of the lattice code. Using $\mathbf{y}_E$ and $\hat{h}_E$ (the estimated channel gain), the most likely transmitted lattice point is computed using the Maximum Likelihood (ML) decoder in \eqref{eq:ML}.

In our experiments, since the codes under consideration have short length and are also small in size, we perform Brute-force ML decoding in \eqref{eq:ML} to recover the secret bits. For decoding conventional codes, \eqref{eq:ML} reduces to the special case $|\mathcal{C}_{j}| = 1, \forall j$ We repeat this decoding procedure for all the codewords in the frame. After the decoding operation, the decoded secret bits from all the frames are collected to reconstruct the \emph{cameraman} image at Eve. Subsequently, Bit Error Rate (BER) is computed by counting the number of bit-positions in which the reconstructed image differs from the original one.

\begin{figure}
\begin{center}
\includegraphics[scale=0.4]{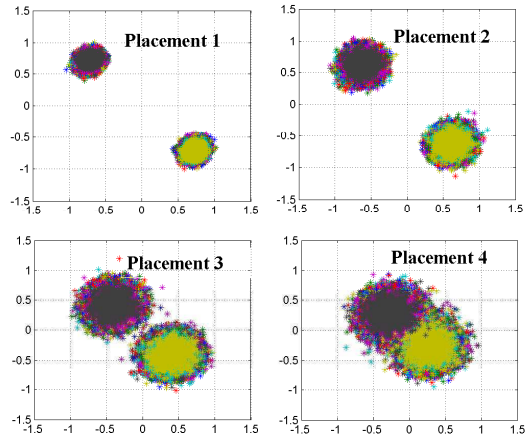}
\end{center}
\vspace{-0.5cm}
\caption{\label{rx_points_conven} The cloud of received points at different placements for conventional coding. The underlying constellation is $\{0, 1\}$, which is shifted and then rotated by $-45$ degrees before transmission.}
\end{figure}

\begin{figure}
\begin{center}
\includegraphics[scale=0.4]{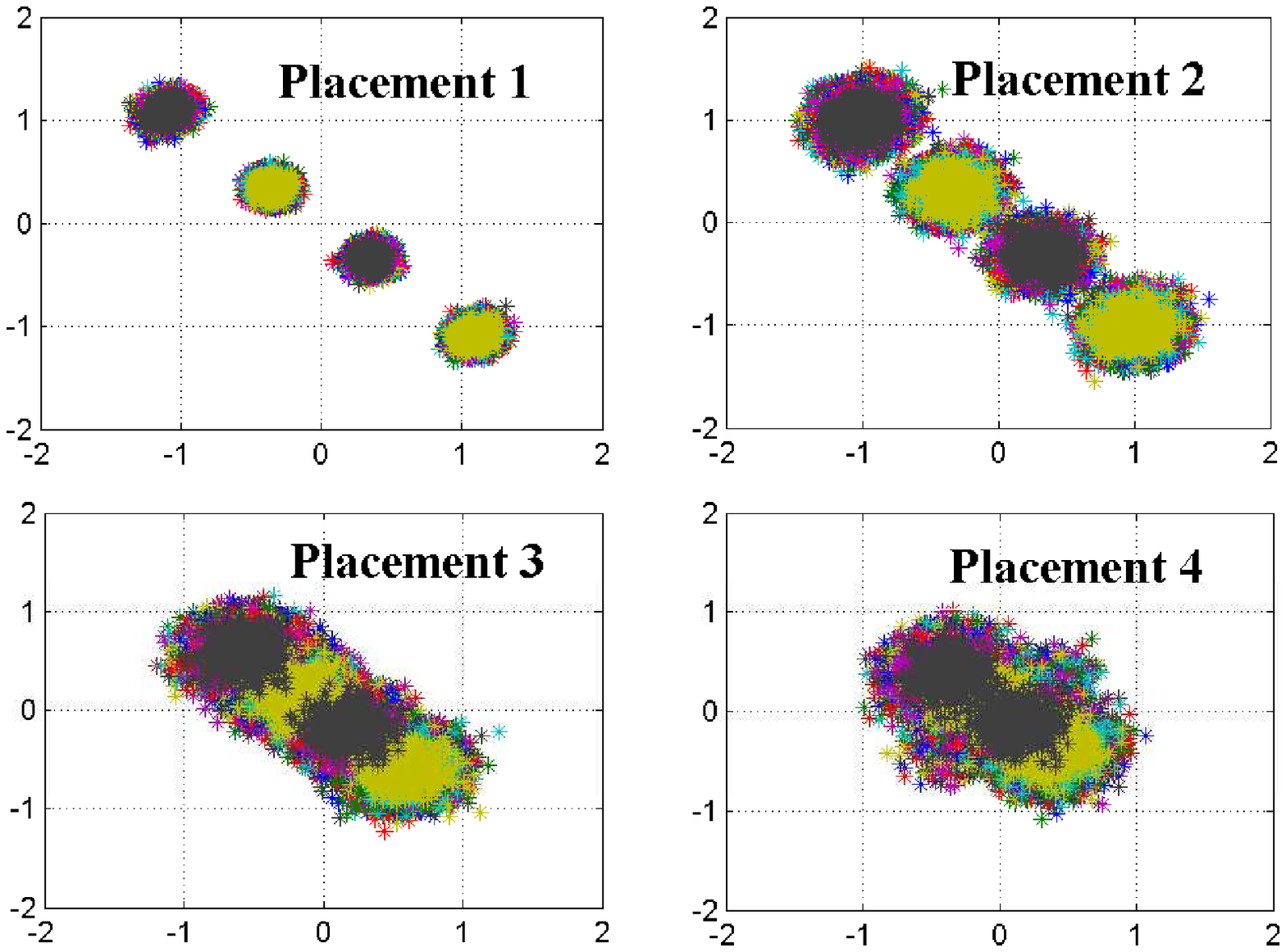}
\end{center}
\vspace{-0.5cm}
\caption{\label{rx_points_1bit} The cloud of received points at different placements for coset encoding with 1 bit confusion. The underlying constellation is $\{0, 1, 2, 3\}$, which is shifted and then rotated by $-45$ degrees before transmission.}
\end{figure}

\begin{figure}
\begin{center}
\includegraphics[scale=0.4]{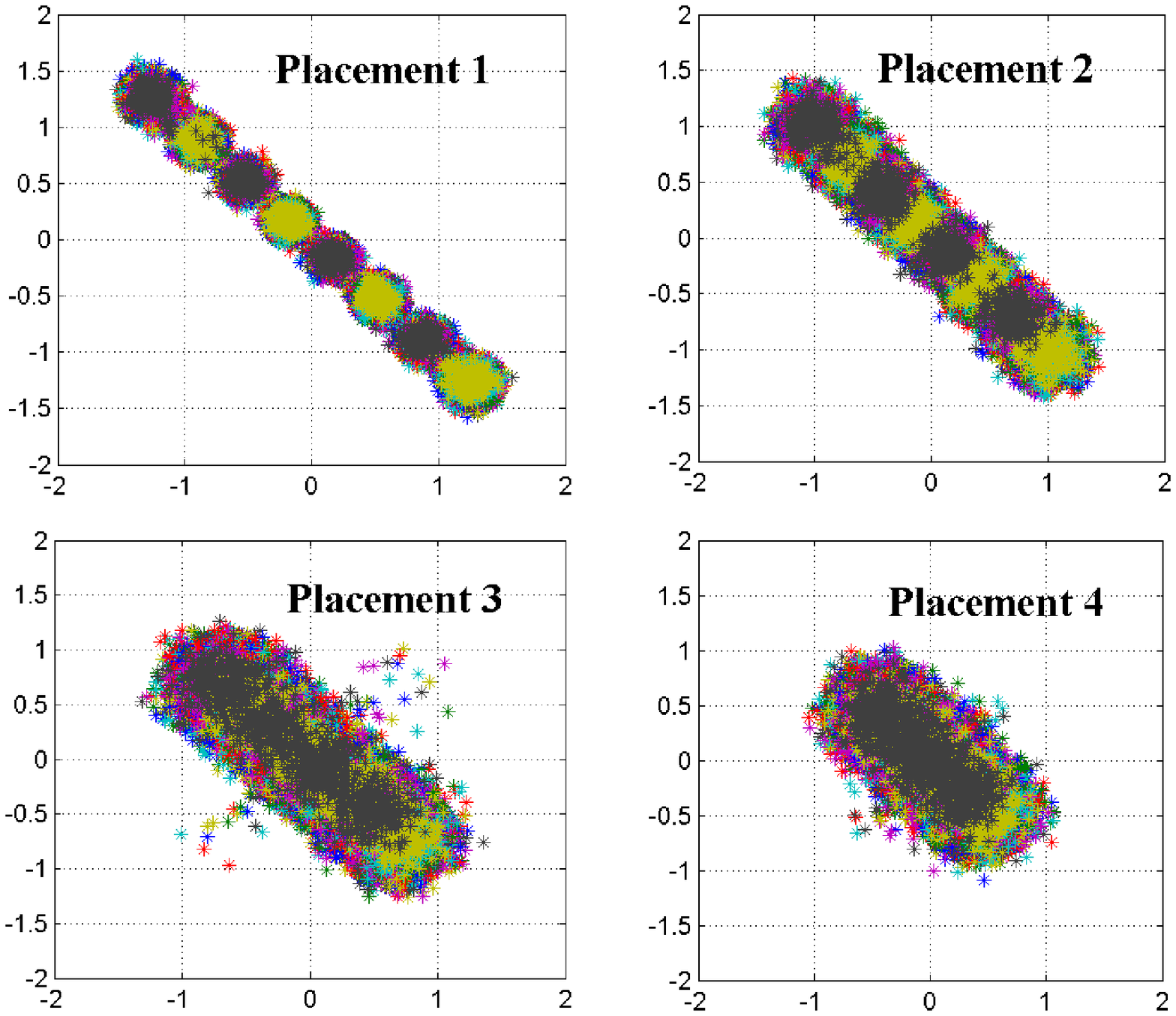}
\end{center}
\vspace{-0.5cm}
\caption{\label{rx_points_2bit} The cloud of received points at different placements for coset encoding with 2 bits confusion. The underlying constellation is $\{0, 1, 2, \ldots, 7\}$, which is shifted and then rotated by $-45$ degrees before transmission.}
\end{figure}

As highlighted earlier, we fix the transmitter USRP position and then vary the receiver USRP position at 5 different locations, referred to as Placement 1, 2, 3, 4 and 5 (in the order of increasing distance). In Fig. \ref{rx_points_conven}, \ref{rx_points_1bit} and \ref{rx_points_2bit}, we plot the received symbols $\{r\}$ (available at the input of the data decoding block in Fig. \ref{coset_decoder_block}) when symbols from the constellation $\{0, 1\}, \{0, 1, 2, 3\}$ and $\{0, 1, 2, \ldots, 7\}$ are transmitted as shown in \eqref{off_set}. Fig. \ref{rx_points_1bit} highlights that for Placement 1, the SNR at Eve is too high to cause degradation in the error performance with both conventional encoding and coset encoding with 1 bit confusion. However, for the same placement, coset encoding with 2 bits confusion can potentially introduce errors since the decision regions overlap. Similarly, for other placements (which corresponds to lower SNR values), it can be seen that coset encoding can potentially result in more errors for Eve. 

\begin{figure}
\begin{center}
\includegraphics[scale=0.4]{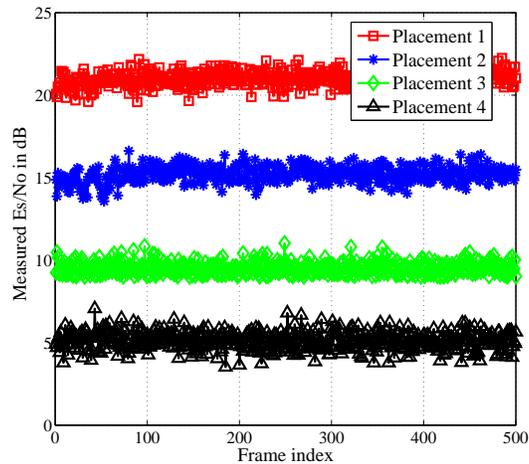}
\end{center}
\vspace{-0.5cm}
\caption{
\label{SNR_meas}
Measured baseband SNR values at four different placements of the receive USRP. The estimated channel gain and the baseband received symbols of each frame are used to measure the SNR values.}
\end{figure}

For the four placements, we compute the corresponding baseband SNR values by transmitting unit energy training sequences. The received symbols are used to compute the SNR as $\frac{\mathbb{E}[|\hat{h}_{E}|^{2}]}{\mathbb{E}[|r|^{2}] - \mathbb{E}[|\hat{h}_{E}|^{2}]}$, where $\mathbb{E}[|\hat{h}_{E}|^{2}]$ indicates the average signal power and $\mathbb{E}[|r|^{2}] - \mathbb{E}[|\hat{h}_{E}|^{2}]$ is indicative of the average noise power. The measured SNR values at different frames are plotted in Fig. \ref{SNR_meas}, which shows that the four placements correspond to the average SNR values of $20.6, 15, 9$ and $4.5$ dB, respectively.

%
%

\begin{figure*}
\begin{center}
\includegraphics[scale=0.6]{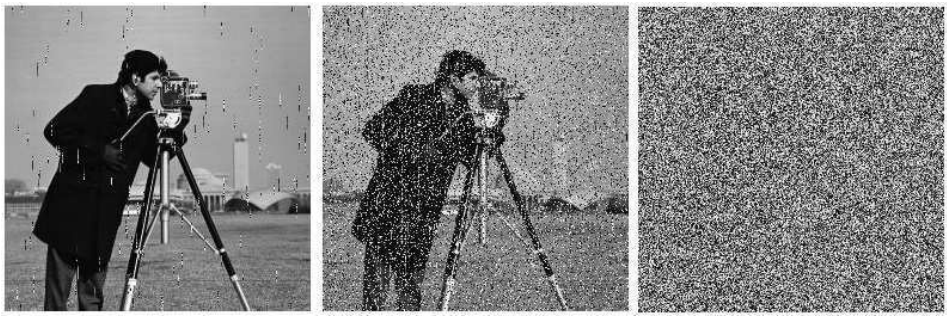}
\end{center}
\vspace{-0.5cm}
\caption{\label{fig:camera_man_rx} Reconstructed images at Eve in Placement 4. From left to right, (a) conventional encoding at Alice, (b) wiretap coding at Alice with one bit confusion, and (c) wiretap coding at Alice with two bits confusion.}
\end{figure*}

\begin{figure}
\begin{center}
\includegraphics[scale=0.4]{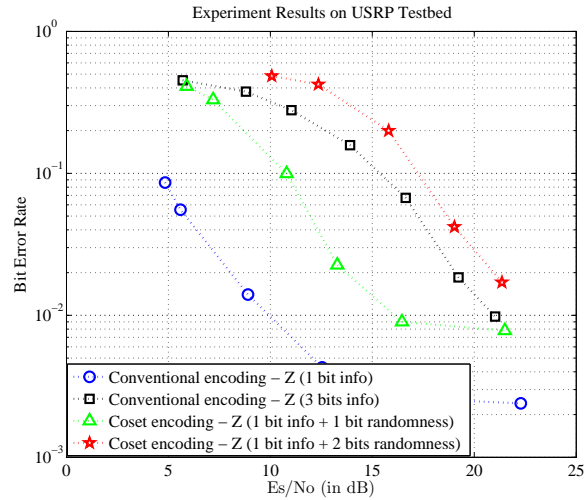}
\end{center}
\vspace{-0.5cm}
\caption{\label{fig:rx_BER_z} Performance comparison between coset and conventional encoding schemes in $\mathbb{Z}$ obtained from the reconstructed images at Eve. The above plots quantify the degradation in the reconstructed image as depicted in Fig. \ref{fig:camera_man_rx}.}
\end{figure}

\begin{figure}
\begin{center}
\includegraphics[scale=0.43]{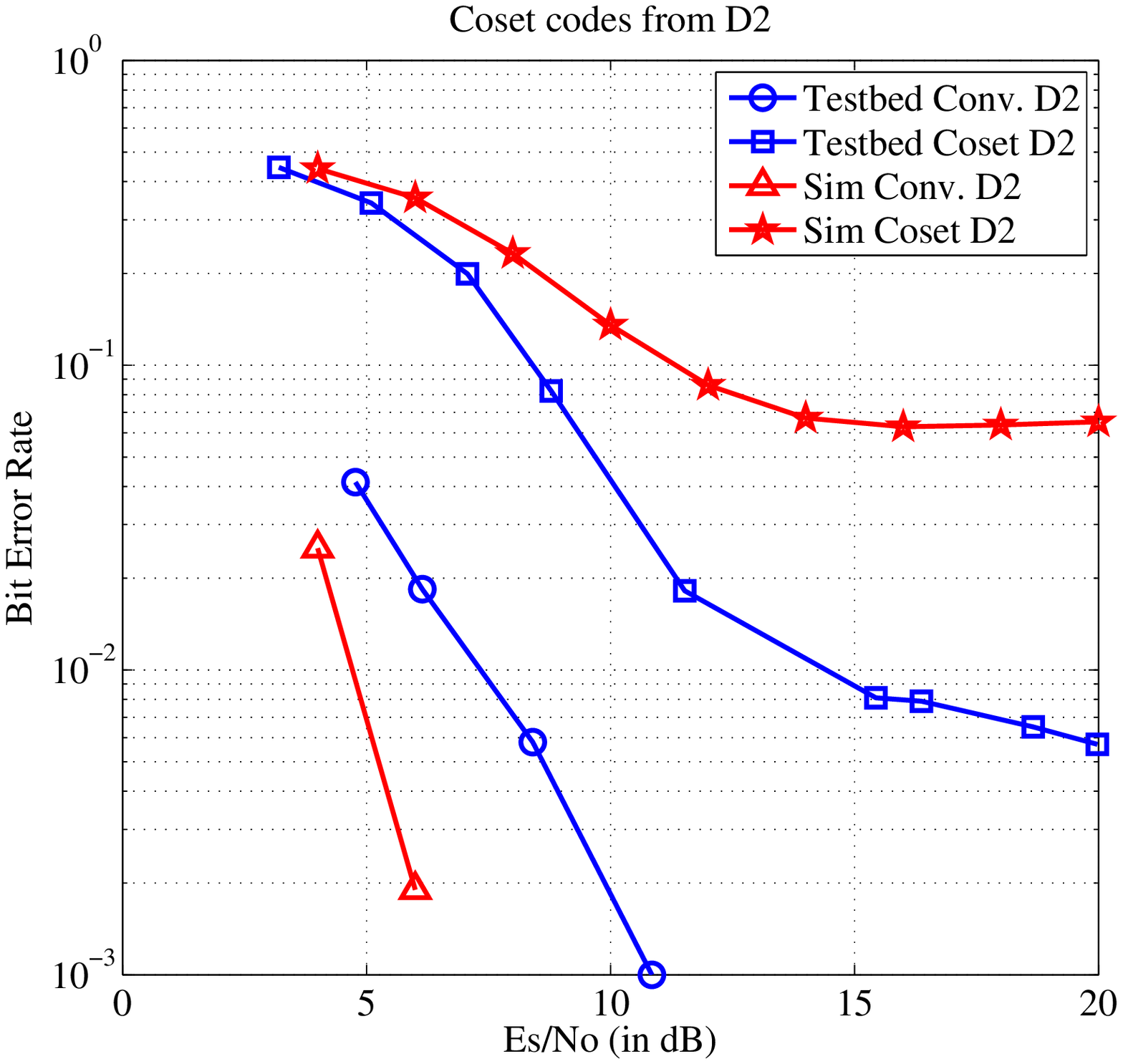}
\end{center}
\vspace{-0.6cm}
\caption{\label{fig:rx_BER_D2} Performance comparison between coset and conventional encoding schemes in $D_{2}$ obtained from experiments on testbed and computer simulations. Both experiments show that coset encoding degrades the error performance at Eve more than conventional encoding.}
\end{figure}

\begin{figure}
\begin{center}
\includegraphics[scale=0.43]{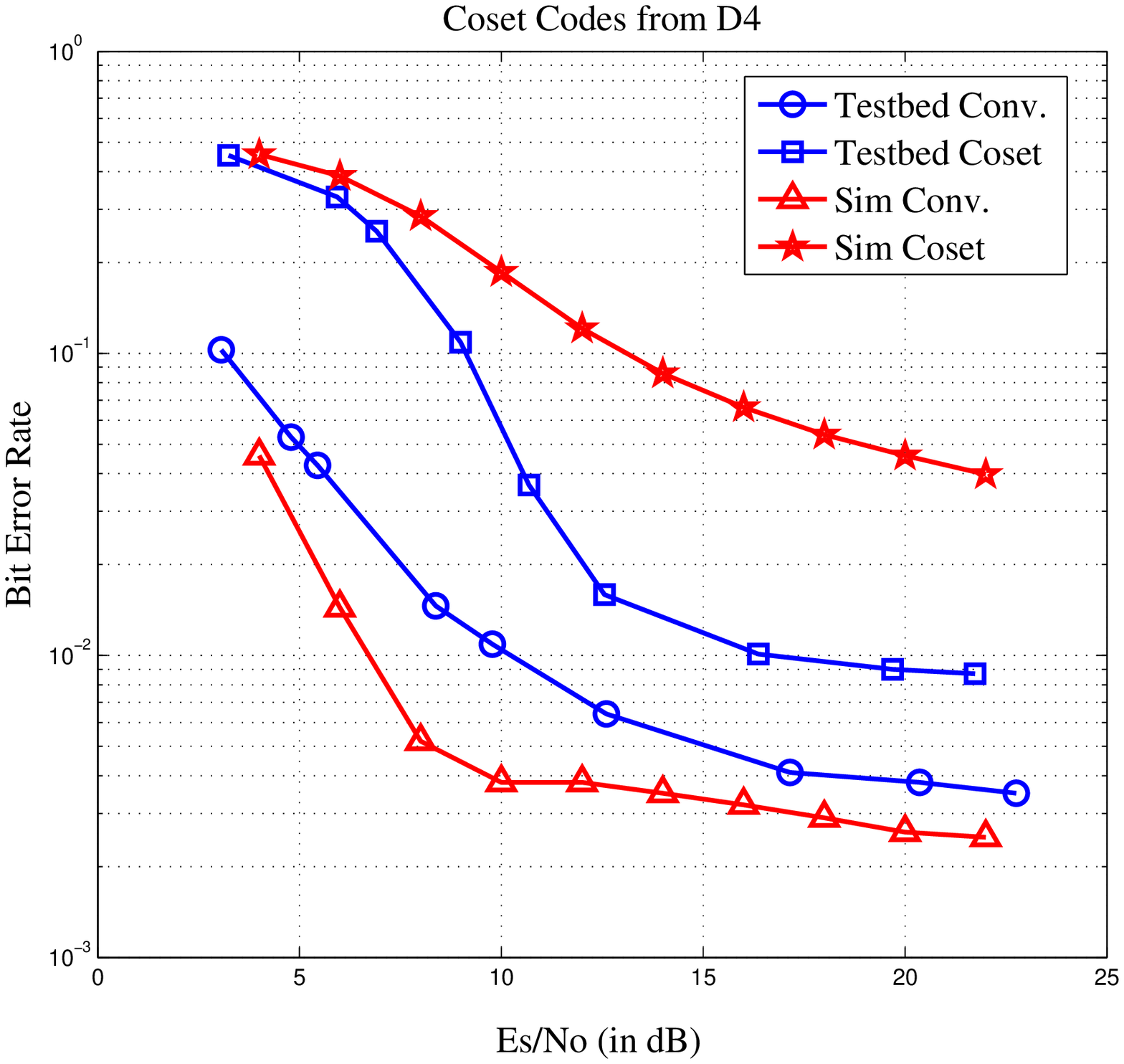}
\end{center}
\vspace{-0.6cm}
\caption{\label{fig:rx_BER_D4} Performance comparison between coset and conventional encoding schemes in $D_{4}$ obtained from experiments on testbed and computer simulations. Both experiments show that coset encoding degrades the error performance at Eve more than conventional encoding.}
\end{figure}

\begin{figure}
\begin{center}
\includegraphics[scale=0.43]{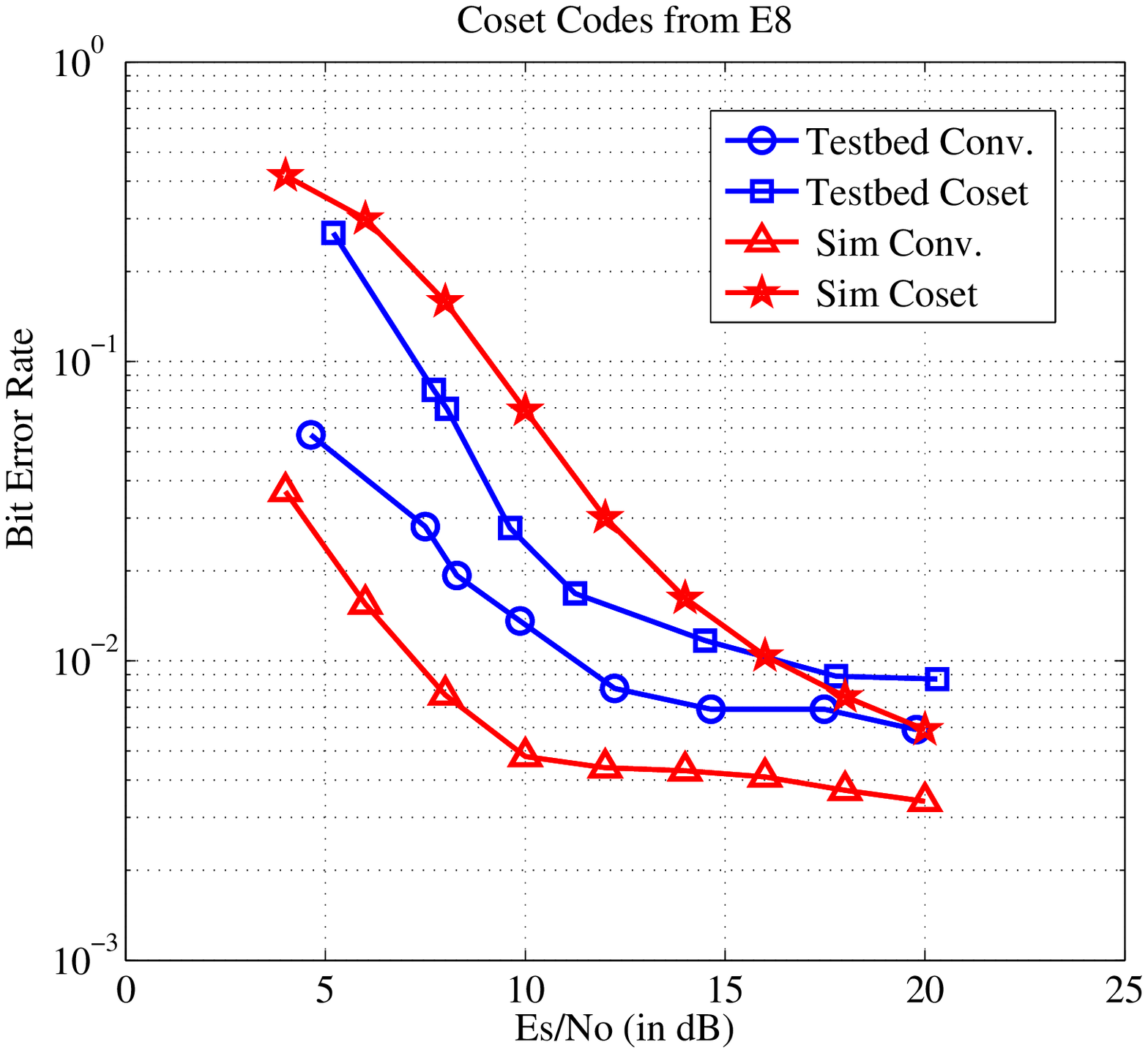}
\end{center}
\vspace{-0.6cm}
\caption{\label{fig:rx_BER_E8} Performance comparison between coset and conventional encoding schemes in $E_{8}$ obtained from experiments on testbed and computer simulations. Both experiments show that coset encoding degrades the error performance at Eve more than conventional encoding.}
\end{figure}

\section{Impact of Coset Coding on the Error Performance at Eve}
\label{sec:exp_results}

The signal transmitted from Alice to Bob is intercepted by Eve, after which it is appropriately sampled and then forwarded to a digital signal processing block to recover the image. Eve is assumed to know everything: the frame structure, modulation scheme, number of bits of confusion, and also the image size. Using the pilot symbols, Eve estimates the beginning of every frame from the sequence of received
symbols and decodes the information bits. For the decoding step, the method employed is either conventional or coset based, depending on the encoding method at Alice (which is presumed known at Eve).

We have conducted the experiments using wiretap codes over $\mathbb{Z}$ with three different encoding schemes: (a) conventional encoding, (b) coset coding with one bit confusion, and (c) coset coding with two bits confusion. The schemes in (a) and (c) are the same as in Section \ref{subsec:ce}, however, the one in (b) is different from (c) as it chooses one bit of randomness instead of two. The reconstructed images at Eve for these cases are displayed in Fig. \ref{fig:camera_man_rx}. From Fig. \ref{fig:camera_man_rx}, it is clear that Eve is able to recover most parts of the image with conventional coding. Although Eve's channel is noisier than that of Bob (by the virtue of longer distance), conventional coding is unable to introduce confusion for Eve at that position. However, it can be seen that coset coding with one bit confusion results in a more distorted image at Eve when compared to conventional coding. Finally, coset encoding with two bits confusion seems to be the best choice for secrecy at Placement 4 as no part of the image is visible. Meanwhile, Bob can correctly recover the image since he is at a distance much closer to Alice, thanks to which, coset coding with two bits confusion does not lead to erroneous reconstruction of the image. In summary, with the knowledge of Eve's distance, Alice is able to decide on the required number of confusion to maintain secrecy. The same behavior is quantified in Fig. \ref{fig:rx_BER_z}, which shows the variation of BER for the above three schemes. 

Other than using basic coset coding over $\mathbb{Z}$ ($L=1$), we also demonstrate the effect of higher dimensional coset lattice codes from $D_{2}$ ($L=2$), $D_{4}$ ($L=4$), and $E_{8}$ ($L = 8$), by using them to transmit the \emph{cameraman image} between the USRP devices. 

\subsection{Coset Codes over Lattices $D_{2}$, $D_{4}$ and $E_{8}$}

For the experiments with $L = 2$, we have carved a lattice code from Construction A of $D_{2}$ given by
\[
D_2=2\ZZ^2+C=(2\ZZ^2+(0,0))\cup(2\ZZ^2+(1,1)).
\]
where $C=\{(0,0),(1,1)\}$ is the repetition code. For conventional lattice coding, one bit of secret is transmitted through the codewords of the repetition code. However, for coset lattice code, we introduce some randomness by choosing non-zero values from $2\ZZ^2$ in Construction A, i.e., we choose $\{(0, 0), (0, 1), (1, 0), (1, 1)\} \subset \mathbb{Z}^{2}$, in order to introduce $2$ bits of randomness per codeword. Therefore, under coset coding, the transmitted points (before scaling and normalization) take values from the set
\begin{equation*}
\{(0, 2), (2, 0), (2, 2), (0, 0)\} + \{(0, 0), (1, 1)\},
\end{equation*}
to carry 2 bits of confusion and 1 bit of secret. Here $+$ is used to denote the direct sum of two sets. From the notations of Section \ref{sec:sod}, the above wiretap lattice code can be written as 
\begin{equation*}
\mathcal{C} = \mathcal{C}_{0} \cup \mathcal{C}_{1},
\end{equation*}
where $\mathcal{C}_{0} = \{(0, 2), (2, 0), (2, 2), (0, 0)\}$ and $\mathcal{C}_{1} = \{(1, 3), (3, 1), (3, 3), (1, 1)\}$.

\noindent Similarly, lattice codes from $D_{4}$ given by
\begin{equation*}
D_{4} = 2\mathbb{Z}^{4} + \mathcal{RM}(1, 2),
\end{equation*}
are also employed in our experiments. Here, $\mathcal{RM}(1, 2) = (n = 4, k = 3, d = 2)$ is the Reed-Muller code of length $4$ and dimension $3$. For this case, conventional lattice encoding will map 3 secret bits to the codewords of $\mathcal{RM}(1, 2)$, while in coset encoding, in addition to the 3 secret bits, some random bits are also mapped from the free part of Construction A, i.e., by forcing $\{0, 1\}^{4} \subset \mathbb{Z}^{4}$. Thus, with coset encoding, 4 bits of randomness are transmitted with 3 bits of secret. Finally, lattice codes from $E_{8}$ given by
\begin{equation*}
E_{8} = 2\mathbb{Z}^{8} + \mathcal{RM}(1, 3),
\end{equation*}
are also employed in our experiments. Here, $\mathcal{RM}(1, 3) = (n = 8, k = 4, d = 4)$ is the Reed-Muller code of length $8$ and dimension $4$. For this case, conventional lattice encoding will map 4 secret bits to the codewords of $\mathcal{RM}(1, 3)$, while in coset encoding, in addition to the 4 secret bits, some random bits are also mapped from the free part of Construction A, i.e., by forcing $\{0, 1\}^{8} \subset \mathbb{Z}^{8}$. Thus, with coset encoding, 8 bits of randomness is transmitted with 4 bits of secret.

The performance comparison between higher-dimensional lattice coset codes and the corresponding conventional lattice codes are presented in Fig. \ref{fig:rx_BER_D2} (for $D_{2}$), Fig. \ref{fig:rx_BER_D4} (for $D_{4}$) and Fig. \ref{fig:rx_BER_E8} (for $E_{8}$), which show that coset encoding offers degraded performance at Eve than that offered by conventional encoding schemes. From Fig. \ref{fig:rx_BER_D2}, we observe that for the location corresponding to 4 dB, coset encoding provides BER of $0.5$ at Eve, whereas conventional encoding offers BER smaller than 10\%. For the rest of the positions (equivalently higher SNR values), Eve gets partial knowledge of the secret even with coset encoding scheme. Therefore, the plots suggests that if Alice has to provide close to $0.5$ BER for other locations, she might have to increase the number of random bits by ensuring that the error performance at Bob does not degrade. 

\subsection{Performance Comparison from Computer Simulations}

Along with the results extracted from the testbed, we also present results obtained from computer simulations in Fig. \ref{fig:rx_BER_D2} to Fig. \ref{fig:rx_BER_E8} (legends in  red), which echo similar observations made from the testbed results. Note that the BER behaviours from the simulations are substantially different from that of the testbed although they too emphasize that coset encoding provides higher confusion than conventional encoding. This difference is attributed to the facts that (i) computer simulations were obtained by simulating an AWGN channel with SNR values corresponding to that of the average SNR measured over all frames, whereas in the testbed experiments, each frame experiences a different SNR level as shown in Fig. \ref{SNR_meas}, and (ii) the entire chain of receiver-side operations pointed in Fig. \ref{coset_decoder_block} is replaced by an additive noise channel in the simulations thereby resulting in different channel model than that in the testbed. 

\section{Discussion}
\label{sec:disc}

In this paper, we have demonstrated the impact of lattice coset codes for introducing confusion at an eavesdropper that experiences a channel that is noisier than that of Bob. In the experiments, we have incorporated the assumption of degradedness by placing Eve farther away from Alice than Bob. As a natural consequence of farther distance, error performance at Eve is worse than that at Bob. In addition, we have shown that adding randomness at the transmitter can further degrade the error performance at Eve. However, additional randomness is carefully added so as not to deteriorate the performance of Bob. One of the main assumptions of this work is that Alice knows the location of Eve (or the equivalent signal-to-noise ratio), and hence, she can decide on the number of random bits. A strong defense to this assumption is the scenario wherein there is physical restriction imposed on the proximity of Eve. This work can be further extended in the following directions: (i) We have conducted experiments using single antenna devices. Advanced physical-layer techniques can be tested on similar testbeds with multiple antenna devices. (ii) We have assumed that the location of the eavesdropper is known to Alice (so as to decide on the amount of randomness to be added). A challenging direction for future work is to detect the presence of an eavesdropper based on the local oscillator leakage \cite{WiR}, and then estimate its distance to decide on the quantum of randomness. Alternatively, bits could be encoded in a hierarchical manner, providing more protection to the most significant ones.


%
\end{document}